**Measurements of Ortho-to-para Nuclear Spin Conversion of $H_2$ on Low-temperature Carbonaceous Grain Analogues: Diamond-like Carbon and Graphite**

Masashi Tsuge[1*], Akira Kouchi[1], and Naoki Watanabe[1]

[1]Institute of Low Temperature Science, Hokkaido University, Sapporo, Hokkaido 060–0819, Japan

*e-mail: tsuge@lowtem.hokudai.ac.jp

**Abstract**

Hydrogen molecules have two nuclear spin isomers: *ortho*-$H_2$ and *para*-$H_2$. The ortho-to-para ratio (OPR) is known to affect chemical evolution as well as gas dynamics in space. Therefore, understanding the mechanism of OPR variation in astrophysical environments is important. In this work, the nuclear spin conversion (NSC) processes of $H_2$ molecules on diamond-like carbon and graphite surfaces are investigated experimentally by employing temperature-programmed desorption and resonance-enhanced multiphoton ionization methods. For the diamond-like carbon surface, the NSC time constants were determined at temperatures of 10–18 K and from $3900 \pm 800$ s at 10 K to $750 \pm 40$ s at 18 K. Similar NSC time constants and temperature dependence were observed for a graphite surface, indicating that bonding motifs ($sp^3$ or $sp^2$ hybridization) have little effect on the NSC rates.

Unified Astronomy Thesaurus concepts: Astrochemistry (75); Molecular clouds (1072); Dense interstellar clouds (371); Interstellar molecules (849); Interstellar dust (836); Laboratory astrophysics (2004);

# 1. Introduction

Molecular hydrogen ($H_2$) has two nuclear spin isomers, ortho and para, in which the two nuclear spins of protons are parallel and antiparallel, respectively. Due to Pauli's exclusion principle, *ortho*- and *para*-$H_2$ molecules in their electronic ground state can only take odd and even rotational quantum numbers ($J$), respectively. Under thermal equilibrium, the ortho-to-para ratio (OPR) is determined by the rotational Boltzmann distribution for odd and even rotational states with spin degeneracies, which are 3 and 1 for *ortho*- and *para*-$H_2$, respectively. The equilibrium OPR value approaches its statistical value of 3 at temperatures above approximately 200 K, whereas it approaches zero at the low temperature limit. At low temperatures, such as 10 K, *ortho*- and *para*-$H_2$ are at their lowest rotational energy levels of $J = 1$ and $J = 0$, respectively, whose energy difference is 14.7 meV (corresponding to ~170 K). Because radiative transition between these states is forbidden in $H_2$ molecules, nuclear spin conversion (NSC) of isolated $H_2$ molecules is extremely slow, with a conversion timescale longer than the age of the universe. Therefore, *ortho*-$H_2$ and *para*-$H_2$ are often considered different species.

The OPR value is an important parameter for the chemistry and physics of $H_2$ in astrophysical environments. It affects the chemical evolution in molecular clouds, such as nitrogen-related chemistry (Dislaire et al. 2012; Faure 2013) and deuterium fractionation. For example, OPR controls the abundance of deuteron donor $H_2D^+$ ions (e.g., Millar 2002) via reaction $H_3^+ + HD \rightleftarrows H_2D^+ + H_2$, where the forward reaction is exothermic by 232 K when the reactant and products are all in the para states. The rate coefficients of the reverse reaction are significantly different for para ($J = 0$) and ortho ($J = 1$) $H_2$ species, and the rate coefficient at 10 K is approximately two orders of magnitude larger for *ortho*-$H_2$ (Gerlich et al. 2002). Therefore, the presence of *ortho*-$H_2$, i.e., higher

OPR, reduces the abundance of $H_2D^+$ and consequently suppresses the formation of deuterated species. From a physical perspective, the OPR affects the gas dynamics of core formation in star-forming regions owing to the different heat capacities of *ortho*- and *para*-$H_2$ (Vaytet et al. 2014).

In the warm gas of molecular shock with a kinetic temperature higher than 300 K, the OPR tends to be lower than the statistical value and ranges from 0.5 to 2 (e.g., Maret et al. 2009), indicating that the OPR is off-equilibrium with the circumstances: non-local thermodynamic equilibrium (non-LTE) OPR. Non-LTE OPR is also commonly observed in photodissociation regions (PDRs) (Fuente et al. 1999; Moutou et al. 1999; Habart et al. 2003; Fleming et al. 2010; Sheffer et al. 2011); proposed mechanisms are either dynamics (advection of low OPR gas into the warm PDR layer; e.g., Fleming et al. 2010), or ortho-to-para NSC on grain surfaces (Le Bourlot et al. 2010; Sheffer et al. 2011; Bron et al. 2016). That is, the OPR in astronomical objects cannot be estimated from their temperatures; therefore, other mechanisms must be identified to determine the OPR of $H_2$. In dense low-temperature clouds, direct astronomical observation of the OPR is impossible for the following reasons: (1) $H_2$ cannot be detected by radio astronomy because *ortho*- and *para*-$H_2$ are populated mostly at their lowest energy levels, i.e., $J = 1$ and 0, respectively, and the radiative pure rotational transitions 3→1 and 2→0 are forbidden due to the absence of permanent dipole moment; (2) $H_2$ is difficult to detect by the ultraviolet absorption (Lyman and Werner bands) because dense clouds are fully opaque in the UV region. Instead, the OPR in dense clouds is often determined indirectly. One method is to deduce the OPR based on the observed abundance of deuterated species, such as $DCO^+$ (Maret & Bergin 2007), $N_2D^+$, and $H_2D^+$ (Pagani et al. 2009), which is affected by the OPR, as mentioned above. Another method employs anomalous rotational absorption lines of formaldehyde (HCOH) that originate from the distinct collision rates with *ortho*- and *para*-H$_2$ (Troscompt et al. 2009).

To understand the meaning of astronomically observed OPRs and deduce the OPR in dense clouds based on astronomical observations, the possible mechanisms for the nuclear spin conversion (NSC) of H$_2$ molecules to change the OPR should be clarified. It has been considered that the NSC of H$_2$ molecules in astrophysical conditions occurs via spin exchange reactions with protons and hydrogen atoms in the gaseous phase. The conversion timescale via reactive collisions with protons or proton-donating ions, such as H$_3^+$, is as slow as $10^5$−$10^7$ yr (Wilgenbus et al. 2000; Flower et al. 2006), and the conversion due to collision with hydrogen atoms is much slower, especially in low-temperature regions, due to the activation barrier (~5000 K). The NSC of H$_2$ by proton exchange at low temperatures has also been studied quantum mechanically (Honvault et al. 2011). Compared with gaseous-phase H$_2$ NSC processes, little is known about H$_2$ NSC processes on astrophysically relevant surfaces; therefore, most astrochemical models of cold cores ignored surface ortho-to-para NSC processes (e.g., in predicting the abundance of deuterated species). Because H$_2$ molecules are predominantly produced on dust grain surfaces via H-H recombination reactions (Wakelam et al. 2017) and gaseous H$_2$ molecules inevitably collide with dust grains, determining how dust grains affect the OPR of H$_2$ is quite important. Watanabe et al. (2010) first demonstrated that the OPR of nascent H$_2$ formed by H-H recombination on a low-temperature surface of amorphous solid water (ASW) is 3. In the last decade, experiments performed by Sugimoto & Fukutani (2011) and Ueta et al. (2016) revealed that H$_2$ NSC occurs on ASW at approximately 10 K on the timescale of $10^2$–$10^4$ s while Chehrouri et al. (2011) deduced conversion timescale of ~$10^4$ s, and the NSC mechanism was discussed theoretically. This experimental finding (i.e., surface NSC processes) has only recently started being incorporated into gas-dust chemical models (Bovino et al. 2017; Furuya et al. 2019).

Since $H_2$ formation starts well before ice mantle formation, Tsuge et al. (2021) investigated the $H_2$ NSC process on an amorphous-$Mg_2SiO_4$ film, which is one of the representative materials of bare grain surfaces. The $H_2$ NSC process on an amorphous-$Mg_2SiO_4$ film was found to be very rapid (timescale of less than $10^3$ s) in a temperature range of 10–18 K. Their numerical simulation showed that the *para*-$H_2$ formation probability per *ortho*-$H_2$ collision with the amorphous-$Mg_2SiO_4$ surface was close to unity at temperatures of 14–18 K. In the present work, we turned our attention to the $H_2$ NSC process on the surface of carbonaceous material, which is another major constituent of bare interstellar grains.

The morphology of carbonaceous grains is considered to be amorphous. An important class of amorphous carbon is tetrahedrally bonded amorphous carbon, ta-C, in which $sp^3$ hybridized carbon dominates (Ong et al. 1995; Robertson 2002; Schultrich 2018; Voevodin & Donley 1996). This class of carbon is often called diamond-like carbon (DLC). Based on the observation of the infrared band characteristic of tetrahedrally bonded carbon toward a young star object, Allamandora et al. (1993) suggested that nanodiamond (or DLC) is ubiquitous in interstellar media. Moreover, the existence of diamond-like material in interstellar media is inferred from the observation of so-called extended red emission, ERE, which extends from 650 to 800 nm range (Chang et al. 2006). In addition, for the purpose of studying the influence of the surface structure ($sp^2$ bonding versus $sp^3$ bonding) and comparing our results with the literature, the $H_2$ NSC process on a graphite surface was also studied in this work.

## 2. Experiment

### *2.1. Preparation and Characterization of DLC film*

A DLC film was prepared by the pulsed laser ablation method as described previously (Tsuge et al. 2019). By preparing it in the vacuum chamber for $H_2$ NSC experiments, surface modification by air was avoided. A rotating carbon target (99.99% graphite; Nilaco Corp.) was irradiated with a focused pulsed laser beam (532 nm wavelength, 8 ns pulse duration, 20 mJ pulse energy), and the laser ablation plume was admitted to deposit on an Al substrate located approximately 40 mm apart from the target at a 45° angle. After irradiation with 6000–9000 pulses, we obtained a 20–30 nm thick DLC film.

Characterization of the DLC sample was performed by using transmission electron microscopy (TEM). A TEM image of a DLC film indicated that the surface was homogeneous. The electron diffraction pattern clearly showed two Debye rings corresponding to 2.12 Å and 1.16 Å, which can be attributed to DLC (Mōri & Namba 1984), whereas a Debye ring originating from graphite at 3.4 Å (Bacon 1951) was not observed. The images are presented in Figure 2 of Tsuge et al. (2019).

### *2.2. Experimental Setup and Procedure*

The experimental setup and procedures have been described previously (Tsuge et al. 2021). The experiments were performed in an ultrahigh vacuum main chamber (base pressure ~$10^{-8}$ Pa). A two-stage differentially pumped $H_2$ molecular beam source was connected to this main chamber. An Al substrate mounted on the cold head of a closed-cycle He refrigerator was placed in the center of the main chamber. The substrate can be cooled to ~6 K, and the temperature, $T_s$, was controlled by using a silicon diode temperature sensor (DT-670, Lake Shore), ceramic heater (MC1020, Sakaguchi E. H VOC), and temperature controller (Model 335, Lake Shore). The temperature fluctuation was less than 0.5 K. For the DLC experiments, the Al substrate was coated with a DLC film as described in Section 2.1, and for the graphite experiments, a commercially obtained highly oriented pyrolytic graphite (HOPG) sheet ( PGCSTM, Panasonic Corp.; Toft-Petersen et al. 2020) was glued onto the Al substrate using silver epoxy. The HOPG sheet was cleaved in air with adhesive tape before the installation to the main chamber.

The $H_2$ NSC processes were investigated with a combination of temperature-programmed desorption (TPD) and resonance-enhanced multiphoton ionization (REMPI) techniques, which is referred to as the TPD-REMPI method. Normal $H_2$ gas (OPR = 3) with a stagnation pressure of 50 kPa was expanded into a vacuum chamber through the 100 μm orifice of a pulsed valve (Series 9, Parker Hannifin) followed by a skimmer, resulting in a pulsed $H_2$ beam. This beam was introduced to the main chamber through an orifice of 1–2 mm. In each measurement, 1000 pulses of $H_2$ beams (300 μs pulse duration, 100 Hz) were deposited onto the substrate. After a certain waiting time ($t_{w.t.}$ of 10–1010 s) after deposition, the substrate was warmed to 55 K at a ramp rate of 20 K min$^{-1}$. The thermally desorbing $H_2$ molecules were ionized by the REMPI method and detected by a linear time-of-flight mass spectrometer equipped with a multichannel plate detector. Laser radiation in the wavelength range 201–203 nm for ionization was obtained from an Nd$^{3+}$:YAG-laser pumped dye laser with subsequent frequency doubling and mixing in nonlinear KDP and BBO crystals. In this wavelength range, *ortho*- and *para*-$H_2$ can be selectively ionized by (2 + 1) REMPI via the $E, F^1(v' = 0, J' = J'') \leftarrow X^1 (v'' = 0, J'' = 0$ or 1) transition.

A pulsed $H_2$ beam was employed instead of a continuous beam because of the following reasons: (1) the duration of $H_2$ deposition can be significantly reduced so that the ortho-to-para conversion during deposition is suppressed; and (2) an undesired increase in background $H_2$ pressure in the main chamber is avoided so that NSC processes on metal surfaces (e.g., on the cold head) are excluded. Nevertheless, deposition of background $H_2$ affected the observed signal intensities, especially for longer $t_{w.t.}$; therefore, blank experiments (i.e., measurements without depositing $H_2$ from the source) were performed to evaluate the contributions from background $H_2$. The substrate was warmed to 55 K prior to each measurement to remove the background $O_2$ molecules deposited on the surface because $O_2$ is known to enhance the NSC process (Chehrouri et al. 2011), and measurements were started just after cooling down to the desired temperature. The deposition of background $H_2O$, which may also affect the NSC process, was suppressed by surrounding the substrate area with a cold copper shroud kept at liquid $N_2$ temperature.

The $H_2$ adsorption site density of our DLC sample was determined to be $(1.5–3) \times 10^{15}$ sites cm$^{-2}$ (Tsuge et al. 2019). The site density was estimated from the $H_2$ saturation dose (i.e., coverage ≈ 1 monolayer) on the DLC sample relative to that on the Al surface, whose adsorption site density was $1.2 \times 10^{15}$ sites cm$^{-2}$ (Wyckoff 1931). The determined adsorption site density was close to the surface atom density of diamond of $\sim 2 \times 10^{15}$ atoms cm$^{-2}$ (Roberts & McKee 1978). The $H_2$ dose was controlled by adjusting the stagnation pressure of $H_2$, opening duration of pulsed valve, and the number of shots. At 10 K, the $H_2$ dose was approximately $5 \times 10^{14}$ molecules cm$^{-2}$, which corresponds to a surface coverage $\theta$ of ~0.2. Although the $\theta$ value was relatively high compared to previous experiments with amorphous-$Mg_2SiO_4$ ($\theta \sim 0.03$), the NSC process due to $H_2$-$H_2$ interactions did not affect the measured NSC rate because it was as slow as 1.90%/h (Silvera 1980). In the case of the graphite sample, the coverage at 10 K was estimated to be 0.2–0.3, assuming a surface atom density of ~4 × $10^{15}$ atoms $cm^{-2}$. At elevated temperatures, $\theta$ should be smaller because of the lower sticking coefficients.

## 3. Results and Discussion

### 3.1. NSC Time Constant on DLC

The experimental results obtained for a DLC sample at surface temperature $T_s$ = 16 K is presented in Figure 1, including a series of TPD-REMPI data measured for $t_{w.t.}$ = 10, 410, and 810 s. For $t_{w.t.}$ = 10 s, $H_2$ deposition was performed at $t$ = 50–60 s and the TPD run was started at $t$ = 70 s. In the case of $J$ = 1 at $t_{w.t.}$ = 10 s, the TPD-REMPI signal started to emerge at 17 K, had a maximum at approximately 23 K, and reached zero near 29 K. The TPD-REMPI signal profile of $J$ = 0 tends to be narrower than that of $J$ = 1 and is slightly shifted to lower temperatures, e.g., at $t_{w.t.}$ = 810 s, the $J$ = 0 signal emerged at 17 K, had a maximum at 22 K, and diminished at 26 K. These differences might reflect a slightly larger binding energy for $J$ = 1 species. The difference has been determined to be approximately 1 meV by simulating TPD spectra, while the distribution of binding energies for $H_2$ and the DLC surface is centered at 30 meV (Tsuge et al. 2019). Figure 1 shows that the $J$ = 0 signal grows and $J$ = 1 decays as $t_{w.t.}$ increases. This behavior clearly indicates that NSC from *ortho*-$H_2$ ($J$ = 1) to *para*-$H_2$ ($J$ = 0) occurs on the DLC surface.

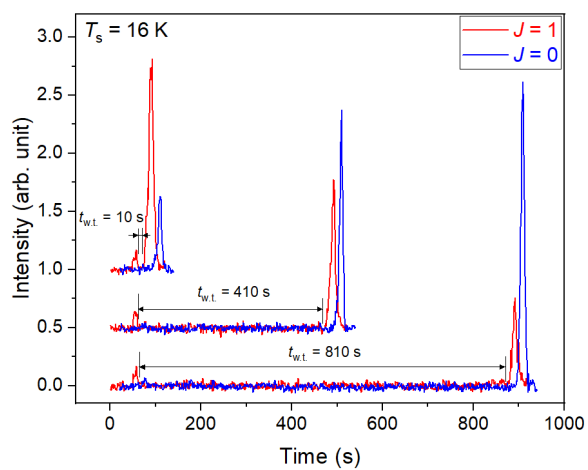

**Figure 1.** Time evolution of the *para*-$H_2$ ($J = 0$, blue) and *ortho*-$H_2$ ($J = 1$, red) populations on a DLC surface at 16 K. From top to bottom, the results for $t_{w.t.}$ = 10, 410, and 810 s are presented. For clarity, the datasets for $t_{w.t.}$ = 10 and 410 s were vertically offset by 1.0 and 0.5, respectively, and the $J = 0$ traces were horizontally offset by 20 s. The weak signals at 50–60 s are due to gaseous $H_2$ from the molecular beam source. The $t_{w.t.}$ is defined as the elapsed time from the termination of $H_2$ deposition and starting time of the TPD run, as indicated by vertical solid lines.

The TPD-REMPI measurements were performed at $T_s$ = 10, 12, 14, 15, 16, and 18 K with $t_{w.t.}$ of 10, 210, 410, 610, 810, and 1010 s. Integrated intensities of $J = 0$ and $J = 1$ signals and their sum are plotted in Figure 2 as a function of the $t_{w.t.}$ value. At surface temperatures up to 16 K, the sum of the $J = 0$ and $J = 1$ signals is constant within the experimental error, as indicated by the dotted lines in Figures 2(a)–2(e), thus indicating that thermal and NSC-induced desorption is negligible. Consequently, the decay of the $J = 1$ signal can be attributed to ortho-to-para conversion. The OPR for $t_{w.t.}$ = 10 s is approximately 3, meaning that the NSC during deposition and the TPD run is negligibly slow.

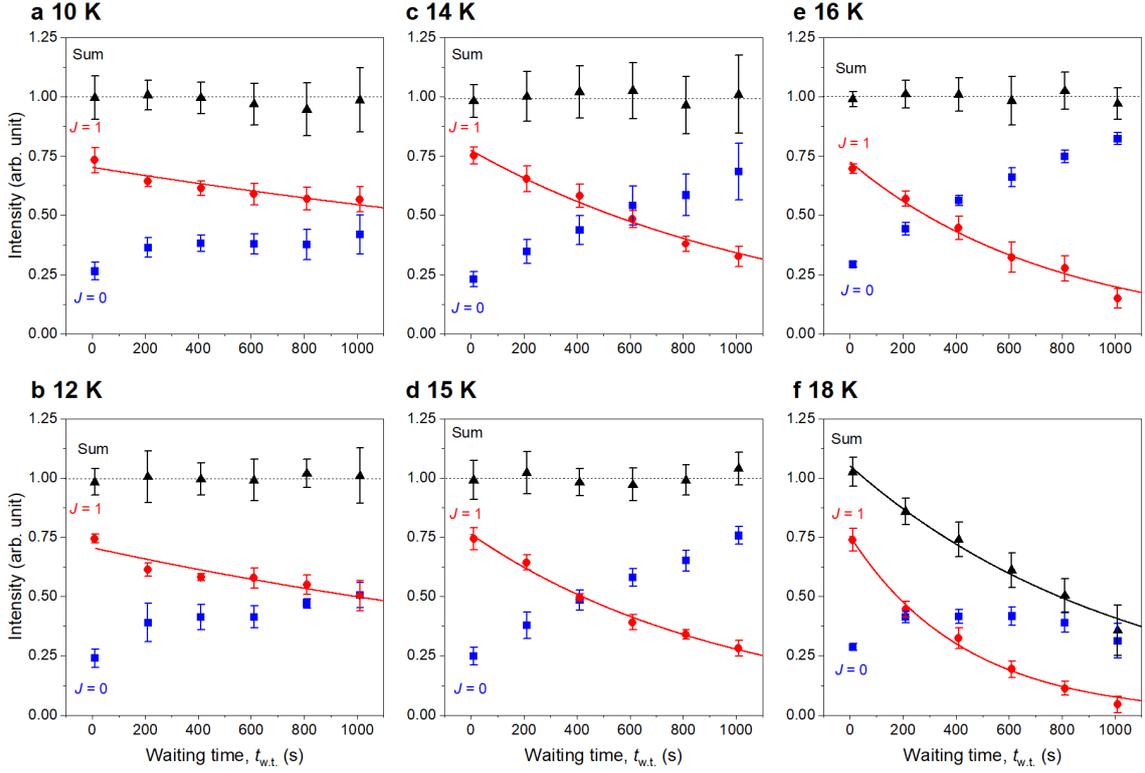

**Figure 2.** Time evolution of the TPD-REMPI intensities for *para*-$H_2$ ($J = 0$, blue squares), *ortho*-$H_2$ ($J = 1$, red circles) and their sum (black triangles) on a DLC surface at temperatures of (a) 10, (b) 12, (c) 14, (d) 15, (e) 16, and (f) 18 K, which are plotted as a function of waiting time ($t_{w.t.}$). The intensities were derived by integrating the TPD-REMPI signal. The derived intensities are scaled so that the average sum becomes unity for $T_s$ = 10, 12, 14, 15, and 16 K and the sum of $t_{w.t.}$ = 10 s becomes unity for $T_s$ = 18 K. The error bars represent the statistical errors among ≥5 measurements. Solid curves are the results of single exponential fitting.

When thermal and NSC-induced desorption processes are negligible, the rate equations for the surface number density of *ortho*-$H_2$ and *para*-$H_2$ are written as follows:

$$\frac{d[ortho\text{-}H_2]}{dt} = -k_{OP}[ortho\text{-}H_2], \qquad (1)$$

$$\frac{d[para\text{-}H_2]}{dt} = k_{OP}[ortho\text{-}H_2], \qquad (2)$$

where $k_{OP}$ represents the ortho-to-para conversion rate. These equations are easily solved to give the following:

$$[ortho\text{-}H_2]_t = [ortho\text{-}H_2]_0 \exp(-k_{OP}t), \qquad (3)$$

$$[para\text{-}H_2]_t = [para\text{-}H_2]_0 + [ortho\text{-}H_2]_0(1 - \exp(-k_{OP}t)), \qquad (4)$$

where $[ortho\text{-}H_2]_t + [para\text{-}H_2]_t = [ortho\text{-}H_2]_0 + [para\text{-}H_2]_0$ = constant. We determined $k_{OP}$ by fitting the decay of the $J = 1$ TPD-REMPI signal by a single exponential function. The $k_{OP}$ values are $(2.5 \pm 0.5) \times 10^{-4}$ s$^{-1}$, $(3.5 \pm 0.7) \times 10^{-4}$ s$^{-1}$, $(3.9 \pm 0.3) \times 10^{-4}$ s$^{-1}$, $(1.01 \pm 0.05) \times 10^{-3}$ s$^{-1}$, and $(1.87 \pm 0.15) \times 10^{-3}$ s$^{-1}$ at 10, 12, 14, 15 and 16 K, respectively, where the errors origites from least-square fitting and statistical errors (shown in Figure 2) were not accounted for. Consequently, the NSC time constants $\tau_{NSC}$ (= $1/k_{OP}$) are 3900 ± 800, 2900 ± 600, 1200 ± 100, 990 ± 50, and 770 ± 60 s at 10, 12, 14, 15, and 16 K, respectively.

For $T_s = 18$ K (Figure 2(f)), the sum of the $J = 0$ and $J = 1$ signals decays as $t_{w.t.}$ increases, indicating that thermal desorption occurs during the waiting time period. In this case, rate equations for the surface number densities of ortho-$H_2$ and para-$H_2$ should be modified as follows:

$$\frac{d[ortho\text{-}H_2]}{dt} = -k_{des}[ortho\text{-}H_2] - k_{OP}[ortho\text{-}H_2], \qquad (5)$$

$$\frac{d[para\text{-}H_2]}{dt} = -k_{des}[para\text{-}H_2] + k_{OP}[ortho\text{-}H_2], \qquad (6)$$

where $k_{des}$ represents the thermal desorption rate. Although the thermal desorption rate for para-$H_2$ is slightly larger than that of ortho-$H_2$, as inferred from the TPD spectra, we adopted a unified rate. By solving Equations (5) and (6), we obtained following equations:

$$[ortho\text{-}H_2]_t = [ortho\text{-}H_2]_0 \exp[-(k_{des} + k_{OP})t], \qquad (7)$$

$$[para\text{-}H_2]_t = ([para\text{-}H_2]_0 + [ortho\text{-}H_2]_0)\exp(-k_{des}t) - [ortho\text{-}H_2]_0\exp[-(k_{des} + k_{OP})t], \qquad (8)$$

$$[H_2]_t = [ortho\text{-}H_2]_t + [para\text{-}H_2]_t = [H_2]_0\exp(-k_{des}t). \qquad (9)$$

Equation (9) indicates that $k_{des}$ is determined by a single exponential fitting of $[H_2]_t$ (the sum of $J = 0$ and $J = 1$ signals), while $k_{des} + k_{OP}$ is determined from $[ortho\text{-}H_2]_t$ according to Equation (7). The least square fitting to the experimental data gave $k_{des} = (9.4 \pm 0.6) \times 10^{-4}$ s$^{-1}$ and $k_{des} + k_{OP} = (2.27 \pm 0.01) \times 10^{-3}$ s$^{-1}$; consequently, we obtained $k_{OP} = (1.33 \pm 0.07) \times 10^{-3}$ s$^{-1}$, which corresponds to $\tau_{NSC} = 750 \pm 40$ s. Taking the difference of 1 meV for the desorption activation energy between *ortho-* and *para-*H$_2$ (Tsuge et al. 2019), the $\tau_{NSC}$ value at 18 K would be as low as 580 s. At higher temperatures (e.g., $T_s = 20$ K), thermal desorption was too fast to accurately determine the NSC time constant.

### *3.2. NSC Time Constant on Graphite*

For H$_2$ NSC on a graphite surface, similar experiments were performed at $T_s = 10$, 12, 14, and 16 K. The integrated intensities of $J = 0$ and $J = 1$ signals and their sum are plotted as a function of $t_{w.t.}$ in Figure 3. At temperatures $T_s = 10$, 12, and 14 K, the sum of the $J = 0$ and $J = 1$ signals were constant, as indicated by dashed lines in Figures 3(a), (b), and (c); therefore, the NSC time constants were determined according to Equation (3). At $T_s = 16$ K, the sum decayed, presumably due to thermal desorption, and the constants were determined according to Equations (7) and (9). The $\tau_{NSC}$ values at $T_s = 10$, 12, 14, and 16 K are determined to be $5400 \pm 800$, $1810 \pm 80$, $920 \pm 30$ s, and $520 \pm 30$ s, respectively. The H$_2$ NSC process at higher temperatures could not be determined due to rapid thermal desorption. The derived $\tau_{NSC}$ values for a graphite surface are similar to those for a DLC surface, indicating that the bonding motifs (i.e., sp$^2$ or sp$^3$ hybridization) may have little effect on the NSC process.

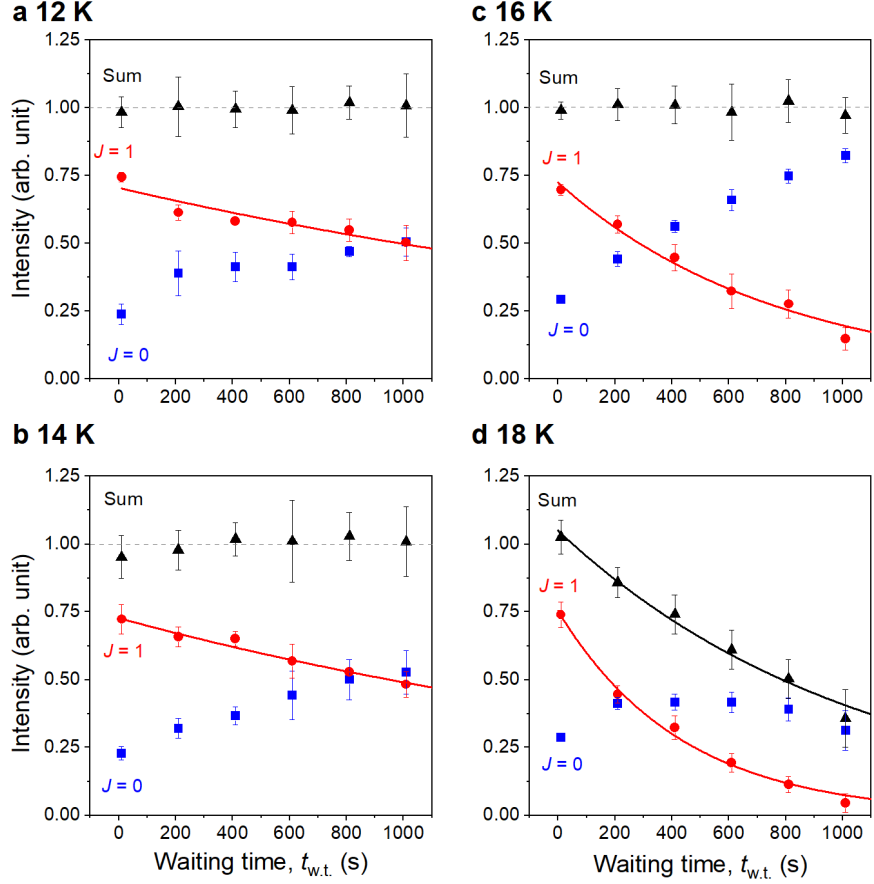

**Figure 3.** Time evolution of the TPD-REMPI intensities for *para*-H$_2$ ($J = 0$, blue squares), *ortho*-H$_2$ ($J = 1$, red circles) and their sum (black triangles) on a graphite surface at temperatures (a) 10, (b) 12, (c) 14, and (d) 16 K, plotted as a function of the waiting time ($t_{w.t.}$). The intensities were derived by integrating the TPD-REMPI signal. The derived intensities are scaled so that the average sum becomes unity for $T_s$ = 10, 12, and 14 K and the sum of $t_{w.t.}$ = 10 s becomes unity for $T_s$ = 16 K. The error bars represent the statistical errors among ≥5 measurements. Solid curves are the results of single exponential fitting.

Here, we compare the derived NSC time constants for a graphite surface with those in the literature. The reported H$_2$ NSC time constants on graphite surfaces are summarized in Table 1. Kubik et al. (1985) used Grafoil samples and Raman spectroscopy to obtain an ortho-to-para conversion rate of 0.40%/h, which corresponds to $\tau_{NSC}$ = 10$^4$ s. Palmer & Willis (1987) reported that they could not observe *ortho*-H$_2$ at 10 K by the electron energy-loss spectroscopy (EELS) and suggested that full conversion occurred within 1 min, which corresponds to $\tau_{NSC} < 20$ s. Yucel et al. (1990) reported that the para-to-ortho NSC of $D_2$ on a Grafoil surface was $1.2 \times 10^3$ s; by using this value and the ratio of ortho-to-para conversion rate of $H_2$ and para-to-ortho conversion rate of $D_2$ reported by Kubik et al. (1985), Bron et al. (2016) deduced that $\tau_{NSC} = 2.5$ s. The $\tau_{NSC}$ determined in this study (5400 s at 10 K) was close to the value $10^4$ s reported by Kubik et al. (1985). However, these values are contradictory to those reported by Palmer & Willis (1987) and Yucel et al. (1990). One possible reason for the contradictory results is that the graphite sample was different. Because magnetic impurities, such as Fe, may be included in the Grafoil sample (Morishita 2011) and paramagnetic centers due to spins of unpaired electrons that occur at defects (Palmer & Willis 1987) should enhance the NSC process, it is desirable to perform several types of measurements with samples prepared with the same procedure. The use of a graphene sheet may be better because the effect of defects will be reduced. Another possibility would be the different degree of adsorption of background $O_2$ molecules, which significantly enhances the ortho-to-para conversion (Chehrouri et al. 2011). In our experiments, we removed adsorbed $O_2$ before each measurement (see Sect. 2.2), whereas it is uncertain whether authors of previous works intentionally avoided the adsorption of background $O_2$.

**Table 1**

Comparison of NSC Time Constants for $H_2$ on the Graphite Surface

| Surface | Method | Temp. (K) | Coverage, $\theta$ | $\tau_{NSC}$ (s) | Reference |
|---|---|---|---|---|---|
| HOPG[a] | TPD-REMPI | 10 | 0.2 | 5400 | This work |
| Grafoil[b] | Raman[c] | <12 | ~1 | $10^4$ | Kubik et al. (1985) |
| HOPG[a] | EELS[d] | 10 | ~1[e] | <20 | Palmer & Willis (1987) |
| Grafoil[a] | Thermal conductivity[f] | 10 | ~1 | 2.5[g] | Yucel et al. (1990) |

**Notes.**

[a] Highly oriented pyrolytic graphite.

[b] A form of exfoliated graphite.

[c] $H_2$ gas was admitted into a low-temperature Grafoil cell in a cryostat. After a certain period, $H_2$ molecules were evaporated by heating the cell and transferred to a gas cell for Raman measurements.

[d] $H_2$ gas was exposed to an HOPG sample kept at 10 K, and electron energy-loss spectroscopy (EELS) was performed in situ.

[e] Palmer & Willis (1985) wrote that coverage was no more than a few layers. Considering that multilayer adsorption is difficult at this temperature, we assumed $\theta$~1.

[f] Experimental procedure is similar to Kubik et al. (1985), and the concentration measurements were performed using a calibrated thermal conductivity cell (Grilly 1953).

[g] Yucel et al. (1990) reported the para-to-ortho conversion time constant of $D_2$ to be $1.2 \times 10^3$ s. Using the ratio of conversion rate for $H_2$ and $D_2$ (Kubik et al. 1985), Bron et al. (2016) deduced the ortho-to-para conversion time constant to be 2.5 s.

### 3.3. NSC Mechanisms

The NSC process involves simultaneous spin conversion (ortho-to-para) and rotational transition. On a surface, the energy associated with rotational transition is dissipated into the surface. A magnetic field will enhance the NSC process (Fukutani & Sugimoto 2013; Ilisca 2021). In our DLC sample, magnetic impurities, such as Fe atoms, and surface sites with unpaired electrons might exist; however, the magnetic effect cannot explain the observed large temperature dependence of $\tau_{NSC}$, i.e., 3900 s at 10 K and 750 s at 18 K. In the present study, we observed the NSC process from $J = 1$ to $J = 0$ rotational energy levels; therefore, the energy gap between these states, 14.7 meV (≈170 K), should be released into the solid. The presence of a large temperature dependence implies that energy dissipation by surface phonons plays a role in the $H_2$ NSC on the DLC surface.

The spin-lattice relaxation process of physisorbed species can be classified into three processes (Scott & Jeffries 1962; Miyamoto et al. 2008): a one-phonon process and two types of two-phonon processes called Raman and Orbach processes. In the one-phonon process, the excess energy associated with the NSC process is dissipated to the surface by excitation of the phonon. The two-phonon process involves the absorption of one phonon and emission of another phonon, and the energy difference between these two phonons corresponds to the excess energy. According to the nature of the intermediate state, namely, real (e.g., $J = 2$ state) or virtual, the two-phonon process is called the Orbach or Raman process, respectively. The former can also be classified as a resonance-Raman process.

The temperature dependence of the NSC rate ($k_{OP} = 1/\tau_{NSC}$) via these processes has been formulated (Scott & Jeffries 1962). The NSC rate due to the one-phonon process will show distinct profiles depending on the energy matching condition between the excess energy and phonon modes of the surface. When the energy matching is satisfactorily good, the one-phonon process will show a temperature dependence, which is described as follows:

$$k_{OP}(T_s) = A \times \coth(\varepsilon/2T_s), \tag{10}$$

where A is a constant and $\varepsilon$ is the energy gap between the $J = 0$ and $J = 1$ states. When there is a mismatch between the excess energy and phonon modes, the process is called a "phonon-limited bottleneck process", and the temperature dependence is described as follows:

$$k_{\text{OP}}(T_s) = B \times T_s^n \ (n \sim 2), \tag{11}$$

where B is a constant. The NSC rate due to the Orbach process presents the following expression:

$$k_{\text{OP}}(T_s) = C \times \exp(-\Delta/T_s), \tag{12}$$

where C is a constant and $\Delta$ represents the energy gap between the initial and intermediate (real) states. This process becomes effective when $\Delta$ is smaller than the Debye temperature ($\theta_D$) of the surface material (Scott & Jeffries 1962). The $\theta_D$ of our DLC film cannot be determined exactly because the fraction of $sp^2$ and $sp^3$ bonds is uncertain. However, it might be in between the $\theta_D$ of diamond ~2200 K and graphite ~1000 K (Yang et al. 2013). On the other hand, the Raman process will show the following power-law dependence when the energy gap ($\varepsilon$) is smaller than $\theta_D$ (Scott & Jeffries 1962) and the surface temperature is sufficiently lower than $\theta_D$ ($T_s < 0.02\theta_D$; van Kranendonk 1954):

$$k_{\text{OP}}(T_s) = D \times T_s^n \ (n = 7 \text{ or } 9), \tag{13}$$

where D is a constant. In the present case, both conditions are met, i.e., $\varepsilon$(~170 K or smaller) $\ll \theta_D$ and $T_s$ (10–18 K) $\approx 0.01\theta_D$. For example, the temperature dependence of the H$_2$ NSC process on the ASW surface has been attributed to the Raman process (Ueta et al. 2016; see Figure 4).

The observed NSC time constants for DLC (this work), amorphous-Mg$_2$SiO$_4$ (Tsuge al. 2021), and ASW (Ueta et al. 2016) are plotted as a function of temperature in Figure 4. Hereafter, we will discuss which model (Equations (10)–(13)) can reproduce the observed dependence for DLC. Among these models, the one-phonon process described by Equation (10) is readily excluded because the calculated rates are almost temperature-independent for the conditions $\delta$ = 170 K and 10 K $\leq T_s \leq$ 18 K. The Orbach process described by Equation (12) with $\Delta$ = 339 K, which corresponds to the energy gap between $J = 1$ and $J = 2$ in the gaseous phase (Silvera 1980), could not reproduce the observed result either. Therefore, the experimental results were fit using the expression $k_{OP}(T_s) = D \times T_s^n$, which describes the temperature dependence for the bottleneck one-phonon process ($n \sim 2$) and two-phonon Raman process ($n = 7$ or $9$). The fitting gave $n = 2.6 \pm 0.5$ for DLC, and the resultant curve is shown in Figure 4. Therefore, we suggest that the $H_2$ NSC process on the DLC surface is dominated by the one-phonon process, which is similar to that on the amorphous-$Mg_2SiO_4$ surface (Tsuge al. 2021).

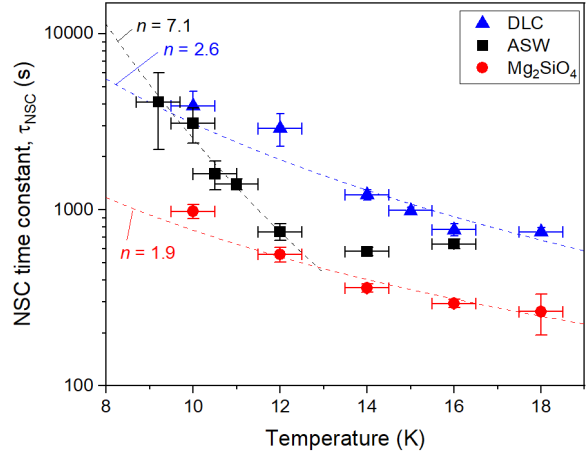

**Figure 4.** Temperature dependence of the NSC time constants for DLC (this work), ASW (Ueta et al. 2016), and amorphous-$Mg_2SiO_4$ (Tsuge et al. 2021) is plotted. Vertical error bars represent the errors originating from the decay curve fitting (Figure 2). Horizontal error bars represent the temperature fluctuations in the experiments. Time constants are summarized in Table 2. Dashed lines represent the results of fitting assuming the $k_{OP}$ (= $1/\tau_{NSC}$) $\propto T^n$ relation; for DLC, ASW, and amorphous-$Mg_2SiO_4$, $n = 2.6 \pm 0.5$, $7.1 \pm 0.6$, and $1.9 \pm 0.3$, respectively, were obtained.

In the one phonon process, the energy of the emitted phonon equals the energy difference between the $J = 1$ and $J = 0$ states. This energy difference is 14.7 meV (~170 K) in the gaseous phase, although it would be smaller due to the hindrance of rotational motion on the surface. On the other hand, carbonaceous solids have phonon modes in the range of 50–180 meV (Lopinski et al. 1996; Mohr et al. 2007), indicating that the energy matching between the energy gap and phonon modes is not so satisfactory that the energy dissipation into the thermal bath is slow. Finally, we note that the one-phonon mechanism would not be the sole mechanism responsible for the NSC of $H_2$ on DLC because the NSC of molecules on the surface is a complicated process (Fukutani & Sugimoto 2013; Ilisca 2021). For example, Ilisca (2018) suggested that the temperature dependence observed for the ASW surface (Ueta et al. 2016) can be reproduced by considering the electromagnetic nature of the process. The proposed mechanism has been described in detail by Ilisca (2018 & 2021); in brief, an excitation of water molecule by spin-orbit coupling leads to Coulomb repulsion between $H_2$ and water molecules and magnetic hyperfine contact between all electrons and protons induces ortho-to-para spin-flip.

### *3.4. Astrophysical Implications*

The $H_2$ NSC time constants on surfaces that are relevant to astrophysical conditions (DLC, amorphous-$Mg_2SiO_4$, and ASW) and on a graphite surface are summarized in Table 2. The time constants on DLC are similar to those on ASW, although the latter shows a different temperature dependence: Equation (13) $n = 7.1 \pm 0.6$ characteristic of the Raman process (Ueta et al. 2016). The NSC process on the amorphous-$Mg_2SiO_4$ surface (Tsuge et al. 2021) was found to be faster by several factors.

**Table 2**

NSC Time Constants Determined for $H_2$ on DLC, Graphite, Amorphous-$Mg_2SiO_4$, and ASW Surfaces at Various Temperatures

| Temperature (K) | NSC Time Constant (s) | | | |
|---|---|---|---|---|
| | DLC[a] | Graphite[a] | $Mg_2SiO_4$[b] | ASW[c] |
| 9.2 | – | | – | 4100 ± 1900 |
| 10 | 3900 ± 800 | 5400 ± 800 | 980 ± 90 | 3100 ± 700 |
| 10.5 | – | | – | 1600 ± 300 |
| 11 | – | | – | 1400 ± 100 |
| 12 | 2900 ± 600 | 1810 ± 80 | 560 ± 60 | 750 ± 80 |
| 14 | 1200 ± 100 | 920 ± 30 | 360 ± 20 | 580 ± 30 |
| 15 | 990 ± 50 | | – | – |
| 16 | 770 ± 60 | 520 ± 30[d] | 290 ± 20 | 640 ± 40 |
| 18 | 750 ± 40[d] | | 260 ± 70 | – |

**Notes.**

[a] This work.

[b] Tsuge et al. (2021).

[c] Ueta et al. (2016).

[d] These values were obtained assuming unified desorption rate for *ortho-* and *para-*$H_2$ (see Sect. 3.1).

In star-forming regions, almost all $H_2$ molecules exist in the gaseous phase rather than on grain surfaces because multilayer binding of $H_2$ to grains is difficult due to the low $H_2$-$H_2$ binding energy (~100 K; Lee et al. 1971), and the number of binding sites (i.e., total surface area of dust grains) is limited ($10^{-4}$ sites per $H_2$). Therefore, it is important to know how the NSC process on dust grains affects the gaseous $H_2$ OPR. The rate of ortho-to-para conversion on dust grain surfaces under astrophysical environments is determined by the (i) collision rate of gaseous *ortho-*$H_2$ with grain surfaces, (ii) sticking probability, (iii) NSC time constant, and (iv) sublimation timescale of adsorbed $H_2$ (Fukutani & Sugimoto 2013; Bovino et al. 2017; Furuya et al. 2019). In previous papers (Furuya et al. 2019; Tsuge et al. 2021), we performed numerical simulations to derive the yield of *para*-$H_2$ per *ortho*-$H_2$ collision with ASW and amorphous-$Mg_2SiO_4$ surfaces based on experimentally derived $k_{OP}$ and binding energy distributions. For the amorphous-$Mg_2SiO_4$ surface, every *ortho*-$H_2$ collision produces approximately one *para*-$H_2$ at temperatures of 14–18 K. At higher and lower temperatures, the efficiency becomes lower due to the limited residence time of $H_2$ on the surface and due to hindrance of *ortho*-$H_2$ adsorption by the preoccupied $H_2$ molecules, respectively. The yield for ASW is approximately 0.5 for the temperature range from 10 to 14 K, and it decreases as the temperature increases and is almost zero at 18 K. Considering that the averaged $H_2$ binding energy on a DLC surface, i.e., 30 meV (~350 K; Tsuge et al. 2019), is slightly smaller than that on ASW (~400 K; He & Vidali 2014) and that the NSC time constants on DLC and ASW are similar (Table 2), the *para*-$H_2$ yield per *ortho*-$H_2$ collision would be highest (~0.5) at temperatures near 10–12 K and become zero below 18 K. To reveal the contribution of NSC processes on dust grain surfaces to gaseous $H_2$ OPR, astronomical simulations of molecular cloud formation, including NSC processes on dust grain surfaces and in the gaseous phase via proton exchange reactions with $H^+$ and $H_3^+$ (Hugo et al. 2009; Honvault et al. 2011), should be performed.

This work was supported by a JSPS Grant-in-Aid for Specially promoted Research (JP17H06087) and partly by a Grant-in-Aid for Scientific Research (B) (JP21H01139).